\documentclass[prd,twocolumn,tightenlines,superscriptaddress,nofootinbib,
showpacs]{revtex4}
\usepackage{amssymb,latexsym}
\usepackage{amsmath,amsbsy,bbm}
\usepackage{epsfig,bm}
\usepackage{graphicx,comment}
\usepackage{color}
\newcommand{\p}{\partial}

\newcommand{\ontopof}[2]{\genfrac{}{}{0pt}{}{#1}{#2}}

\begin{document} 

\title{Loop-Cluster Simulation of the $t$-$J$ Model
on the Honeycomb Lattice}

\author{F.-J.~Jiang}
\email[]{fjjiang@itp.unibe.ch}
\affiliation{Institute for Theoretical Physics, Bern University, Sidlerstrasse 5, CH-3012 Bern, Switzerland}
 
\author{F.~K\"ampfer}
\email[]{fkampfer@itp.unibe.ch}
\affiliation{Institute for Theoretical Physics, Bern University, Sidlerstrasse 5, CH-3012 Bern, Switzerland}

\author{M.~Nyfeler}
\email[]{nyfeler@itp.unibe.ch}
\affiliation{Institute for Theoretical Physics, Bern University, Sidlerstrasse 5, CH-3012 Bern, Switzerland}

\author{U.-J.~Wiese}
\email[]{wiese@itp.unibe.ch}
\affiliation{Institute for Theoretical Physics, Bern University, Sidlerstrasse 5, CH-3012 Bern, Switzerland}

\vspace{-1cm}

\begin{abstract}
Inspired by the lattice structure of the unhydrated variant of the   
superconducting material Na$_x$CoO$_2 \cdot$yH$_2$O at $ x = \frac{1}{3}$, 
we study the $t$-$J$ model on a honeycomb lattice by using an efficient 
loop-cluster algorithm. The low-energy physics of the undoped system and of 
the single hole sector is described by a systematic low-energy
effective field theory. The staggered magnetization per spin 
$\widetilde{{\cal M}}_s = 0.2688(3)$, 
the spin stiffness $\rho_s = 0.102(2) J$, 
the spin wave velocity $c= 1.297(16) J a$, and the kinetic mass $M'$ of a hole are 
obtained by fitting the numerical Monte Carlo data to the effective theory 
predictions.

\end{abstract}

\maketitle

\section{Introduction}
\label{intro}
Since the discovery of high-temperature superconductivity 
in cuprate materials \cite{Bednorz86}, 
the Hubbard and $t$-$J$ models have been of central importance 
in strongly correlated electron systems. 
However, due to their strong coupling, a systematic analytic treatment
of these models is currently not available. Similarly, a severe sign
problem away from half-filling prevents us from understanding 
these systems quantitatively by reliable Monte Carlo calculations. 
Despite these
difficulties, much effort has been devoted to understanding the
properties of $t$-$J$-type models on the square lattice. Although some
controversial results have been obtained, various studies including exact
diagonalization \cite{Eder96,Lee97}, series expansion \cite{Hamer98}, and
Monte Carlo simulations \cite{Brunner00,Mishchenko01} enable us to understand
the hole-dynamics quantitatively, at least to some extent. 
In particular, these studies all
obtained minima of the single-hole dispersion relation at lattice momenta
$(\pm\frac{\pi}{2a},\pm\frac{\pi}{2a})$ in the Brillouin zone of the square
lattice which is in agreement with experimental results~\cite{Wel95,LaR97,Kim98}. 

A reliable and order-by-order exact way to investigate the
low-energy physics of lightly doped antiferromagnets is provided by a 
systematic
low-energy effective field theory. The physics of the undoped systems is 
quantitatively described by magnon chiral perturbation theory~\cite{Cha89,Neu89, 
Fis89,Has90,Has91},
while the interactions of magnons and holes are described by a low-energy
effective theory for hole-doped antiferromagnets~
\cite{Kampfer:2005ba,Brugger:2006dz}. 
Predictions of the effective
theory only depend on a small number of low-energy constants which can be
determined from either experiments or Monte Carlo data. Thus, the 
use of low-energy effective theories together with reliable Monte Carlo
simulations provides an unbiased approach to studying the low-energy
physics of these systems. In particular, using the loop-cluster 
algorithm \cite{Eve93}, the low-energy parameters of the spin $1/2$
Heisenberg model have been determined with very high precision 
\cite{Wie94,Bea96}. Indeed, thanks to the combination of very efficient 
Monte Carlo simulations with the systematic low-energy effective field 
theory, undoped antiferromagnets on the square lattice like 
La$_2$CuO$_4$ and Sr$_2$CuO$_2$Cl$_2$ are among the quantitatively
best understood condensed matter systems.    

In addition to the cuprates, another superconducting 
material, Na$_x$CoO$_2 \cdot$yH$_2$O, has drawn a lot of attention both
theoretically and experimentally. Unfortunately, due to the fact that the underlying 
lattice geometry of the spin $1/2$ cobalt sites in these materials 
is triangular --- which leads to strong geometric frustration --- 
a first principles Monte Carlo study is impossible in practice. 
Nevertheless, the spin- and charge-ordering tendencies observed and studied in 
\cite{Motrunich04,Zheng04,Watanabe05} may suggest that 
at filling $x = \frac{1}{3}$, the unhydrated parent compound Na$_x$CoO$_2$
can be described by the $t$-$J$ model on a half-filled honeycomb lattice which allows 
one to simulate the system efficiently with the loop-cluster 
algorithm. 

Another system on the honeycomb lattice that has been investigated with great
vigor is graphene --- a single sheet of graphite (see \cite{Net07} 
for a detailed review).
As a consequence of the geometry
of the honeycomb lattice, the low-energy excitations of graphene are massless
Dirac fermions. If some variants of graphene exists at stronger coupling, one 
eventually expects a phase transition separating graphene's unbroken phase 
from a strong coupling antiferromagnetic phase in which the $SU(2)_s$ spin 
symmetry is spontaneously broken to $U(1)_s$. The low-energy effective theory 
of the unbroken phase and of the critical point has been constructed in 
\cite{Her06}.    
    
Motivated by possible applications to Na$_x$CoO$_2$, we investigate the
spin $1/2$ Heisenberg model as well as the $t$-$J$ model on the honeycomb 
lattice by using the quantum Monte Carlo method. 
Just as in the square lattice case, the long-distance
physics of these models is described quantitatively by a systematic 
low-energy effective field theory. At low energies, the Heisenberg model 
on a bipartite lattice
is described by magnon chiral perturbation theory, and, accordingly, the $t$-$J$ model
is described by a low-energy effective field 
theory for magnons and holes. Based on the same method that has been used in
the square lattice case, we have constructed the leading-order terms in the action
of a systematic low-energy effective field theory for magnons and holes on the
honeycomb lattice. In this paper, we determine the corresponding leading-order
low-energy constants, namely the staggered magnetization per spin 
$\widetilde{{\cal M}}_s$, the spin stiffness $\rho_s$,
the spin wave velocity $c$, and the kinetic mass $M'$ of a doped hole by 
fitting the Monte Carlo data to the effective field theory predictions. 
 
The rest of this paper is organized as follows. In section \ref{micro}, 
we introduce 
the relevant microscopic models as well as corresponding observables. Section \ref{lowen} 
reviews the low-energy effective theory for magnons, and section \ref{deter} is devoted 
to the Monte Carlo determination of the corresponding low-energy parameters.
The single-hole physics is investigated in section \ref{singl}, and the 
effective theory for holes and magnons is discussed in section \ref{effec}. Finally,
section \ref{concl} contains our conclusions.

\section{Microscopic Models and Corresponding Observables}
\label{micro}
In this section we introduce the Hamiltonians of the microscopic 
$t$-$J$ model and the Heisenberg model
as well as some relevant observables.
The $t$-$J$ model is defined by the Hamilton operator
\begin{eqnarray}
H = P \big\{- t \sum_{\langle x y \rangle}
(c_x^\dagger c_{y} + c_{y}^\dagger c_x) +
J \sum_{\langle x y \rangle} \vec S_x \cdot \vec S_{y}\big\} P. 
\end{eqnarray}
Here $c^{\dagger}_x$ and $c_x$ are fermion creation
and annihilation operators at a site $x$ with
\begin{equation}
c_x = \left(\begin{array}{c} c_{x \uparrow} \\ c_{x \downarrow}
\end{array} \right),
\end{equation}
whose components obey standard anticommutation relations.
In terms of the Pauli matrices $\vec \sigma$ the local spin operator 
at a site $x$ is given by
\begin{equation}
\vec S_x= c_x^\dagger\, \frac{\vec \sigma}{2} \, c_x\,.
\end{equation}
The projection operator $P$ restricts the Hilbert space by eliminating doubly
occupied sites. 
Hence the $t$-$J$ model allows empty or singly occupied sites
only. 
The hopping of fermions is controlled by the parameter $t$, while $J > 0$ is
the antiferromagnetic exchange coupling between neighboring spins. At
half-filling, the $t$-$J$ model reduces to the Heisenberg model with the Hamiltonian
\begin{eqnarray}
H = J \sum_{\langle x y \rangle} \vec S_x \cdot \vec S_{y}\,.
\end{eqnarray}

The honeycomb lattice with periodic spatial boundary conditions implemented in 
our simulations is depicted in figure 1. 
The dashed rectangle in figure 1, which contains $4$ spins, is the elementary
cell for building a periodic honeycomb lattice covering a rectangular area.  
For instance, the honeycomb lattice shown in figure 1 contains 3 $\times$ 3 
elementary cells. The lattice
spacing $a$ is 
the distance between two neighboring sites. The honeycomb lattice is not a
Bravais lattice. 
Instead it consists of two triangular Bravais sub-lattices $A$ and $B$
(depicted by solid and open circles in figure 1). As a consequence, 
the momentum space of the honeycomb lattice is a doubly-covered Brillouin 
zone of the two triangular sub-lattices (depicted in figure 2).

A physical quantity of central interest is the staggered susceptibility
which is given by
\begin{eqnarray}
\label{defstagg}
\chi_s &=& \frac{1}{L_{1} L_{2}} 
\int_0^\beta dt \ \langle M^3_s(0) M^3_s(t) \rangle \nonumber \\
&=& \frac{1}{L_{1} L_{2}} \int_0^\beta dt \ \frac{1}{Z} 
\mbox{Tr}[M^3_s(0) M^3_s(t) \exp(- \beta H)].
\end{eqnarray}
Here $\beta$ is the inverse temperature, $L_{1}$ and $L_{2}$ are the
spatial box sizes
in the $x_1$- and $x_2$-direction, respectively, and 
\begin{equation}
Z = \mbox{Tr}\exp(- \beta H)
\end{equation} 
is the partition function. 
The staggered magnetization order parameter $\vec{M}_s$ is defined by
\begin{equation}
\vec M_s = \sum_x (-1)^{x} \vec S_x\,.
\end{equation}
Here $(-1)^{x} = 1$ on the $A$- and $(-1)^{x} = -1$ on the $B$-sub-lattice,
respectively.
Another relevant quantity is the uniform susceptibility which is given by
\begin{eqnarray}
\label{defuniform}
\chi_u &=& \frac{1}{L_{1} L_{2}} \int_0^\beta dt \ 
\langle M^3(0) M^3(t) \rangle \nonumber \\
&=& \frac{1}{L_{1} L_{2}} \int_0^\beta dt \ \frac{1}{Z} \mbox{Tr}[M^3(0) M^3(t)
\exp(- \beta H)].
\end{eqnarray}
Here
\begin{equation} 
\vec{M} = \sum_x \vec S_x
\end{equation} 
is the uniform magnetization. 
Both $\chi_s$ and $\chi_u$ can be measured very efficiently with the 
loop-cluster algorithm using improved estimators \cite{Wie94}. In particular, 
in the multi-cluster version of the algorithm the staggered susceptibility
is given in terms of the cluster sizes $|{\cal C}|$ (which have the dimension
of time), i.e.
\begin{equation}
\chi_s = \frac{1}{4 \beta L^2} \left\langle \sum_{\cal C} |{\cal C}|^2 
\right\rangle\,.
\end{equation}
Similarly, the uniform susceptibility
\begin{equation}
\chi_u = \frac{\beta}{4 L^2} \left\langle W_t^2 \right\rangle =
\frac{\beta}{4 L^2} \left\langle \sum_{\cal C} W_t({\cal C})^2 
\right\rangle
\end{equation}
is given in terms of the temporal winding number
$W_t = \sum_{\cal C} W_t({\cal C})$ which is the sum of winding numbers
$W_t({\cal C})$ of the loop-clusters ${\cal C}$ around the Euclidean time 
direction. Similarly, the spatial winding numbers are defined
by $W_i = \sum_{\cal C} W_i({\cal C})$ with $i \in \{1,2\}$. 

\begin{figure}
\label{bchoneycomb}
\begin{center}
\includegraphics[width=0.45\textwidth]{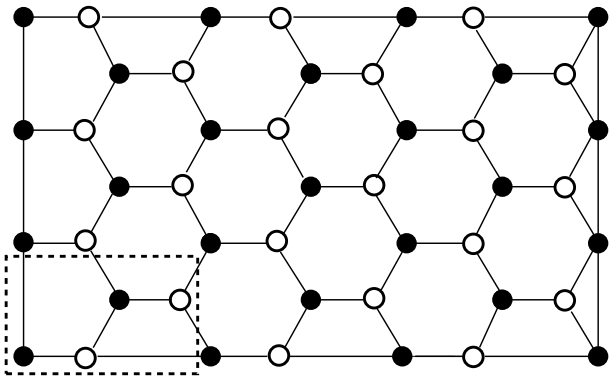}
\caption{\it The periodic honeycomb lattice consisting of two triangular
  sub-lattices A and B, which are depicted by solid and open 
circles, respectively. The dashed rectangle is an elementary cell
for building a periodic honeycomb lattice covering a rectangular area.}
\end{center}
\end{figure}

\begin{figure}
\label{brillouin}
\begin{center}
\includegraphics[width=0.45\textwidth]{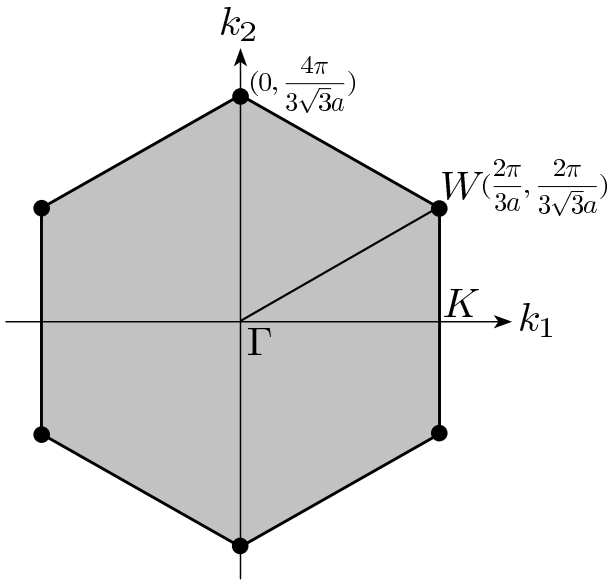}
\caption{\it The momentum space of a honeycomb lattice, which is a 
doubly-covered Brillouin zone dual to the two triangular sub-lattices
A and B.}
\end{center}
\end{figure}

\section{Low-Energy Effective Theory for Magnons}
\label{lowen}
Due to the spontaneous breaking of the $SU(2)_s$ spin symmetry down to
its $U(1)_s$ subgroup, the low-energy physics of antiferromagnets is governed by
two massless Goldstone bosons, the antiferromagnetic 
spin waves or magnons.
The description of the low-energy magnon physics
by an effective theory was pioneered by Chakravarty, Halperin, and Nelson 
in \cite{Cha89}. In analogy to chiral perturbation theory for the 
pseudo-Goldstone pions in QCD, a systematic low-energy 
effective field theory for magnons was developed in 
\cite{Neu89,Fis89,Has90,Has91}. The staggered magnetization of an 
antiferromagnet can be described by a unit-vector field $\vec{e}(x)$
in the coset space $SU(2)_s/U(1)_s = S^2$, i.e. 
\begin{equation}
\vec e(x) = \big(e_1(x),e_2(x),e_3(x)\big), \qquad \vec e(x)^2 = 1\,.
\end{equation}
Here $x = (x_1,x_2,t)$ denotes a point in (2+1)-dimensional space-time. 
To leading order, the Euclidean magnon low-energy effective action takes 
the form
\begin{equation}
S[\vec e\,] = \int d^2x \ dt \ \frac{\rho_s}{2} 
\left(\p_i \vec e \cdot \p_i \vec e +
\frac{1}{c^2} \p_t \vec e \cdot \p_t \vec e\right)\,,
\end{equation}
where the index $i \in \{1,2\}$ labels the two spatial directions and
$t$ refers to the Euclidean time-direction. The parameter $\rho_s$ is the spin 
stiffness and $c$ is the spin wave velocity. At low energies the 
antiferromagnet has a relativistic spectrum. 

Using the above Euclidean action, detailed calculations of a variety of physical quantities including 
the next-to-next-to-leading 
order contributions have been carried out in
\cite{Has93}. Here we only quote the results that are relevant 
for our study, namely the
finite-temperature and finite-volume effects of the staggered and uniform
susceptibilities, as well as results on the rotor spectrum of the
antiferromagnet in a finite volume.   
The aspect ratio of a spatially quadratic space-time box with
$L_{1} = L_{2} = L$ is characterized 
by $l = (\beta c /L )^{1/3}\,,$
with which one distinguishes cubical space-time volumes with 
$\beta c \approx L$ from cylindrical ones with
$\beta c \gg L$. 
In the cubical regime the volume- and temperature-dependence of the staggered 
susceptibility is given by
\begin{eqnarray}
\label{chiscube}
\chi_s &=& \frac{{\cal M}_s^2 L^2 \beta}{3} 
\left\{1 + 2 \frac{c}{\rho_s L l} \beta_1(l)
  \right. \nonumber \\
&+&\left.\left(\frac{c}{\rho_s L
      l}\right)^2 \left[\beta_1(l)^2 
+ 3 \beta_2(l)\right] + O\left(\frac{1}{L^3}\right) \right\},
\end{eqnarray}
where ${\cal M}_s$ is the staggered magnetization density. The uniform 
susceptibility takes the form
\begin{eqnarray}
\label{chiucube}
\chi_u &=& \frac{2 \rho_s}{3 c^2} 
\left\{1 + \frac{1}{3} \frac{c}{\rho_s L l} \widetilde\beta_1(l)
  \right. \nonumber \\
&+&\left.\frac{1}{3} \left(\frac{c}{\rho_s L l}\right)^2
\left[\widetilde\beta_2(l) - \frac{1}{3} \widetilde\beta_1(l)^2 - 6 \psi(l)
\right] \right. \nonumber \\
&+& \left. O\left(\frac{1}{L^3}\right) \right\}.
\end{eqnarray}
The functions $\beta_i(l)$, $\widetilde\beta_i(l)$, and $\psi(l)$, 
which only depend on $l$, are shape 
coefficients of the space-time box defined in \cite{Has93}.
In the very low temperature limit, one enters the cylindrical regime of
space-time volumes with $\beta c \gg L$. In this case, the staggered
magnitization vector $\vec{M}_s$ acts as a quantum rotor and, 
correspondingly, the low-energy end of the spectrum takes the form
\begin{equation}
E_{S} = \frac{S(S+1)}{2\Theta}.
\end{equation}
Here $S \in \{0,1,2,...\}$ is the spin and $\Theta$ is the moment of inertia
of
the quantum rotor which is given by \cite{Has93}
\begin{equation}
\Theta = \frac{\rho_s L^2}{c^2}\left[ 1 + \frac{3.900265c}{4\pi\rho_s L} + O\left(\frac{1}{L^2}\right)\right].
\end{equation}
The partition function of the (2$S$+1)-fold degenerate rotor spectrum is
given by
\begin{equation}
Z = \sum_{S=0}^{\infty}(2S+1)\exp\left(-\beta E_S\right).
\end{equation}
The probability distribution of the uniform magnetization $M^3 = S^3$ is then
given by
\begin{equation}
\label{hist}
p(M^3) = \frac{1}{Z}\sum_{ S\ge |M^3|} \exp\left(-\beta E_S\right).
\end{equation}

\section{Determination of the Low-Energy Parameters of the Undoped System}
\label{deter}
In order to determine the low-energy constants ${\cal M}_{s}$, $\rho_{s}$, and 
$c$, we have performed numerical simulations of the Heisenberg model on
the honeycomb lattice with up to $4680$ spins in the cubical and
cylindrical regimes. The cubical regime is determined by the condition 
$\,\langle \sum_C W_{1}(C)^2 \,\rangle \approx \langle\, \sum_C W_{2}(C)^2 \,\rangle \
\approx \langle\, \sum_C W_{t}(C)^2\, \rangle$ (which implies $\beta c \approx
L$). The chiral perturbation 
theory predictions for $\chi_s$ and $\chi_u$ in Eqs.(\ref{chiscube}) and 
(\ref{chiucube}) are derived for a (2+1)-dimensional box with 
equal extent in the two spatial directions (which we refer to as a 
square-shaped area). Since it is not possible to consider the honeycomb 
geometry on an exactly
square-shaped area, our simulations are done on almost square-shaped 
rectangles. To be more
precise, the lattices used in our simulations deviate from a
perfect square-shaped area by less than $0.4$ percent. 
We have performed an interpolation on some of our 
data to the exactly square-shaped area and find agreement between the 
fits of the interpolated data and the raw data. The inclusion 
of ${\cal O}(1/L^3)$ corrections in the fits leads to consistent results 
as well.
Instead of considering the staggered magnetization density ${\cal M}_s$
of Eq.(\ref{chiscube}), we choose to quote the staggered magnetization per 
spin $\widetilde{{\cal M}}_s$, which is related to ${\cal M}_s$ by
\begin{equation}
\widetilde{{\cal M}}_s = \frac{3\sqrt{3}}{4}{\cal M}_s a^2. 
\end{equation} 

\begin{figure}
\begin{center}
\vspace{-0.15cm}
\epsfig{file=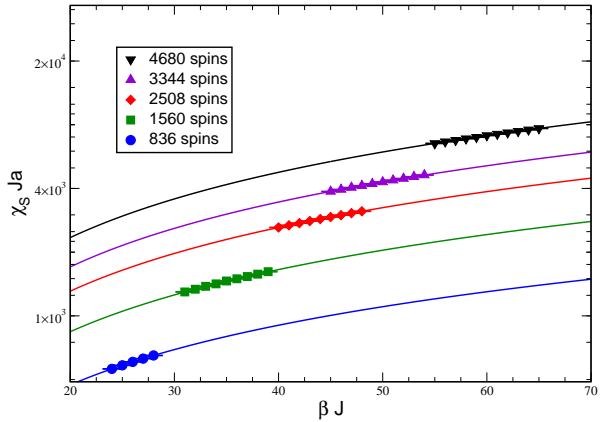,angle=-90,width=8.0cm} \vskip-0.5cm
\end{center}
\caption{\it Fit of the finite-size and finite-temperature effects of the
staggered susceptibility $\chi_s$ to results of the effective 
theory in the cubical regime.}
\label{chiralfit1}
\end{figure}

\begin{figure}
\begin{center}
\vspace{-0.15cm}
\epsfig{file=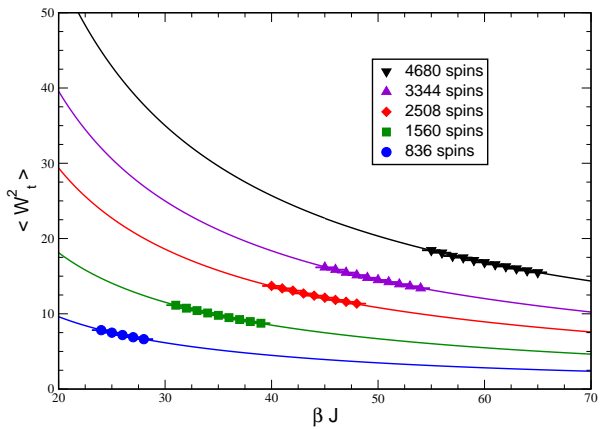,angle=-90,width=8.0cm} \vskip-0.5cm
\end{center}
\caption{\it Fit of the finite-size and finite-temperature effects of 
the temporal winding number squared
$\langle W_t^2 \rangle$ to the results of the effective theory in 
the cubical regime.}
\label{chiralfit}
\end{figure}

Some numerical data from our simulations are listed in table 1. By fitting 
$\chi_s$ and $\chi_u$ simultaneously to Eq.(\ref{chiscube}) and 
Eq.(\ref{chiucube}), we find
\begin{eqnarray}
\widetilde{{\cal M}}_s = 0.2688(3),\, \rho_s = 0.102(2) J,\, 
c = 1.297(16) J a  
\end{eqnarray}
with $\chi^2/{\text{d.o.f.}} \approx 1.05$ (see figures~\ref{chiralfit1} 
and \ref{chiralfit}). The low-energy constants $\rho_s$ and $c$ 
are determined with high accuracy (at the percent level). 
The error of $\widetilde{{\cal M}}_s$ is even at the permille level.
The value of $c$ obtained here is consistent with 
the one of a spinwave expansion study \cite{Mat94}.
The above value of $\widetilde{{\cal M}}_s$ is larger
than the one of a previous spinwave expansion \cite{Wei91} but consistent 
with 
that of a series expansion study \cite{Oit92} (within the comparably large 
$4$ percent error of that study). It is only slightly larger than the value 
obtained in a previous Monte 
Carlo calculation $\widetilde{{\cal M}}_s = 0.2677(6)$
\cite{Cas05}.
We want to point out that our results are obtained by fitting more than $80$ 
numerical data points to two analytic predictions with only 3 unknown 
parameters. 
If $\widetilde{{\cal M}}_s$ is fixed to $0.2677$, 
the quality of our fit downgrades to  $\chi^2/{\text{d.o.f.}} \approx 3.0$.  
The reduction of 
$\widetilde{{\cal M}}_s = 0.2688(3)$ and $\rho_s = 0.102(2)J$ on the honeycomb 
lattice compared to those 
on the square lattice ($\widetilde{{\cal M}}_s = 0.3074(4)$, 
$\rho_s = 0.186(4)J$ \cite{Wie94, Bea96}) indicates larger quantum 
fluctuations on the honeycomb lattice. This is expected since the 
coordination number of the honeycomb lattice is smaller than the one of 
the square lattice.

\begin{figure}
\begin{center}
\vspace{-0.15cm}
\epsfig{file=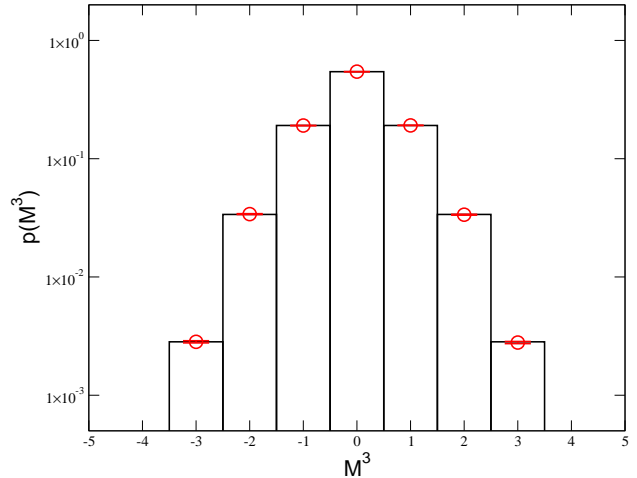,angle=-90,width=9.0cm} \vskip-0.5cm
\end{center}
\caption{\it Comparison of the effective theory prediction 
for the probability distribution $p(M^3)$ of Eq.(\ref{hist}) 
with Monte Carlo data on an $N_1 \times N_2 = 11 \times 19$ honeycomb lattice 
with $N_{\text{Spin}} = 836$ at $\beta J$ = 60. The open circles are the 
Monte Carlo data while the histogram is the effective theory prediction.}
\label{histgram}
\end{figure}

Having determined the values of the low-energy parameters 
$\widetilde{{\cal M}}_s$, $\rho_s$, and $c$ from the cubical space-time
volume regime, we can test the effective theory in the cylindrical regime.
Figure \ref{histgram} shows a comparison of the effective theory prediction 
for the
probability distribution $p(M^3)$ of Eq.(\ref{hist}) with Monte Carlo 
data. The observed excellent agreement --- which does not involve any 
adjustable parameters --- confirms the quantitative correctness of
the effective theory.  

\begin{table}
\label{data}
\begin{center}
\begin{tabular}{|c|c|c|c|c|c|}
\hline 
$\beta J$&
$N_1$&
$N_2$&
$N_{\text{Spin}}$&
$\chi_s J a$&
$\langle W_t^2 \rangle$\tabularnewline
\hline
\hline 
$24$&
$11$&
$19$&
$836$&
$575.14(82)$&
$7.828(15)$\tabularnewline
\hline 
$25$&
$11$&
$19$&
$836$&
$597.58(85)$&
$7.494(15)$\tabularnewline
\hline 
$26$&
$11$&
$19$&
$836$&
$620.91(85)$&
$7.177(15)$\tabularnewline
\hline 
$34$&
$15$&
$26$&
$1560$&
$1450(3)$&
$10.113(20)$\tabularnewline
\hline 
$35$&
$15$&
$26$&
$1560$&
$1496(3)$&
$9.797(21)$\tabularnewline
\hline 
$36$&
$15$&
$26$&
$1560$&
$1532(3)$&
$9.491(22)$\tabularnewline
\hline 
$44$&
$19$&
$33$&
$2508$&
$2936(5)$&
$12.411(25)$\tabularnewline
\hline 
$45$&
$19$&
$33$&
$2508$&
$3001(5)$&
$12.145(25)$\tabularnewline
\hline 
$46$&
$19$&
$33$&
$2508$&
$3061(5)$&
$11.848(26)$\tabularnewline
\hline 
$48$&
$22$&
$38$&
$3344$&
$4220(6)$&
$15.137(28)$\tabularnewline
\hline 
$49$&
$22$&
$38$&
$3344$&
$4308(7)$&
$14.796(26)$\tabularnewline
\hline 
$50$&
$22$&
$38$&
$3344$&
$4392(7)$&
$14.495(28)$\tabularnewline
\hline
$59$&
$26$&
$45$&
$4680$&
$7151(11)$&
$17.123(29)$\tabularnewline
\hline 
$60$&
$26$&
$45$&
$4680$&
$7286(11)$&
$16.838(29)$\tabularnewline
\hline 
$61$&
$26$&
$45$&
$4680$&
$7401(12)$&
$16.557(31)$\tabularnewline
\hline
\end{tabular}
\end{center}
\caption{\it Some numerical data for the staggered susceptibility $\chi_s$ and the
temporal winding number squared $\langle W_t^2 \rangle$ obtained with the
loop-cluster algorithm. ${\text{N}}_1$ and ${\text{N}}_2$ count the number of
copies of elementary rectangles in the $1$- and $2$-direction and 
${\text{N}}_{\text{Spin}} = 4 {\text{N}}_1 {\text{N}}_2$ is the corresponding number of spins.}
\end{table}

\section{Single Hole Dispersion Relation}
\label{singl}
\begin{figure}
\begin{center}
\vspace{-0.15cm}
\epsfig{file=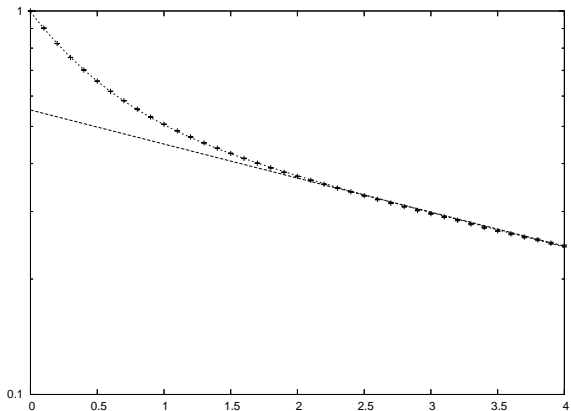,angle=-90,width=8.00cm} \vskip-0.5cm
\end{center}
\caption{\it Correlation function $G^{AA}(k,t)$ between AA sub-lattices with Fourier momentum
  $k = (\frac{2\pi}{3a},\frac{2\pi}{3\sqrt{3}a})$. The bottom line is the result of a single-exponential fit
  while the top line is obtained from a double-exponential fit.} 
\label{correlation_fit}
\end{figure}

\begin{figure}
\begin{center}
\vspace{-0.15cm}
\epsfig{file=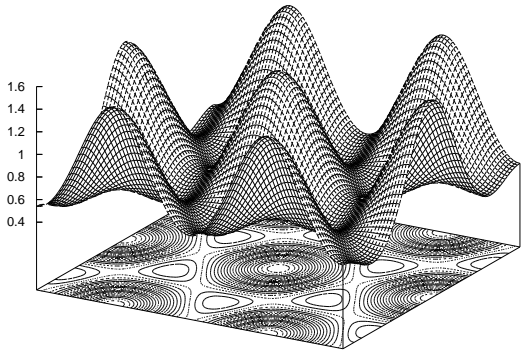,angle=-90,width=8.8cm} \vskip-0.5cm
\end{center}
\caption{\it The dispersion relation $E_{h}(k)/t$ for a single hole in an 
antiferromagnet on the honeycomb lattice.}
\label{single_landscape}
\end{figure}

The physics of a single hole on the honeycomb lattice was studied theoretically
in \cite{Luscher04} using exact diagonalization, series expansion,
and self-consistent Born approximation. Here we use first principle 
Monte Carlo simulations to quantitatively investigate the single-hole 
dispersion relation and quasiparticle weight. 
To achieve this goal, we have implemented a technique similar to the ones
used in \cite{Brunner00,Mishchenko01} to simulate the one-hole sector of 
the $t$-$J$ model. To calculate the fermion two-point 
functions in momentum space, one should keep in mind that the Brillouin zone 
of the honeycomb lattice is doubly covered. The two covers are dual to the 
two triangular Bravais sub-lattices $A$ and $B$. Therefore, one needs to 
distinguish the correlators between $AA$, $AB$, $BA$, and $BB$ sub-lattices. 
The correlation function between AA sub-lattices with momentum $k$ takes the form
\begin{eqnarray}
\label{correlation}
G^{AA}(k,t) &=& \frac{1}{Z}\sum_{x,y \in
  A}{\text{Tr}}[c^{\dagger}_{x}(0)c_{y}(t)\exp(-\beta H)] \nonumber \\
  &&\times\exp(-ik(x-y)) \nonumber \\ 
&\sim& \sum_{n=1}^{\infty} Z_n(k) \exp\left(-(E_{n}(k)-E_{0})t\right),
\end{eqnarray}
where $E_0$ is the ground state energy of the half-filled system, and 
\begin{equation}
Z_n(k) = |\langle 0| \sum_{x \in A}c_x \exp(ikx)|n\rangle |^2. 
\end{equation}
The factor $Z_{1}(k)$ is known as the quasiparticle weight.
In deriving Eq.(\ref{correlation}), we have inserted a complete set
of energy eigenstates ${\bf{1}} = \sum_{n}| n \rangle\langle n |$ in
the single-hole sector 
and taken the limit $\beta \rightarrow \infty$ in the final step.
The fermion energy
\begin{equation} 
E_{h}(k) = E_{1}(k)-E_{0} 
\end{equation}
corresponding to the momentum $k$ can be extracted 
by fitting the data to a single- or a double-exponential. 
The correlation function between $AA$ sub-lattices with momentum 
$k = (\frac{2\pi}{3a},\frac{2\pi}{3\sqrt{3}a})$ depicted in
figure \ref{correlation_fit} is obtained on a honeycomb lattice with 3456 spins
and $J/t = 2.0$. 
A single-exponential fit  
yields $E_{h}(k) = 0.207(9)t$ while a double-exponential fit results in 
$E_{h}(k) = 0.201(5)t$. The two fits yield consistent
results. In the same way, we determine the one-hole dispersion relation 
from the AA correlator for all momenta $k$. 
The single-hole dispersion relation in figure \ref{single_landscape} is 
obtained with the same parameters as in figure \ref{correlation_fit}. The figure shows that the hole pockets 
are located at $(\pm\frac{2\pi}{3a},\pm\frac{2\pi}{3\sqrt{3} a})$ and 
$(0,\pm \frac{4\pi}{3\sqrt{3} a})$ in the Brillouin zone. The position of the 
hole pockets agrees with the position of the Dirac cones obtained from the 
free fermion theory on the honeycomb lattice which is relevant for graphene. 

\section{Effective Field Theory for Holes and Magnons}
\label{effec}
\begin{figure}
\vskip-2.5cm
\begin{center}
\epsfig{file=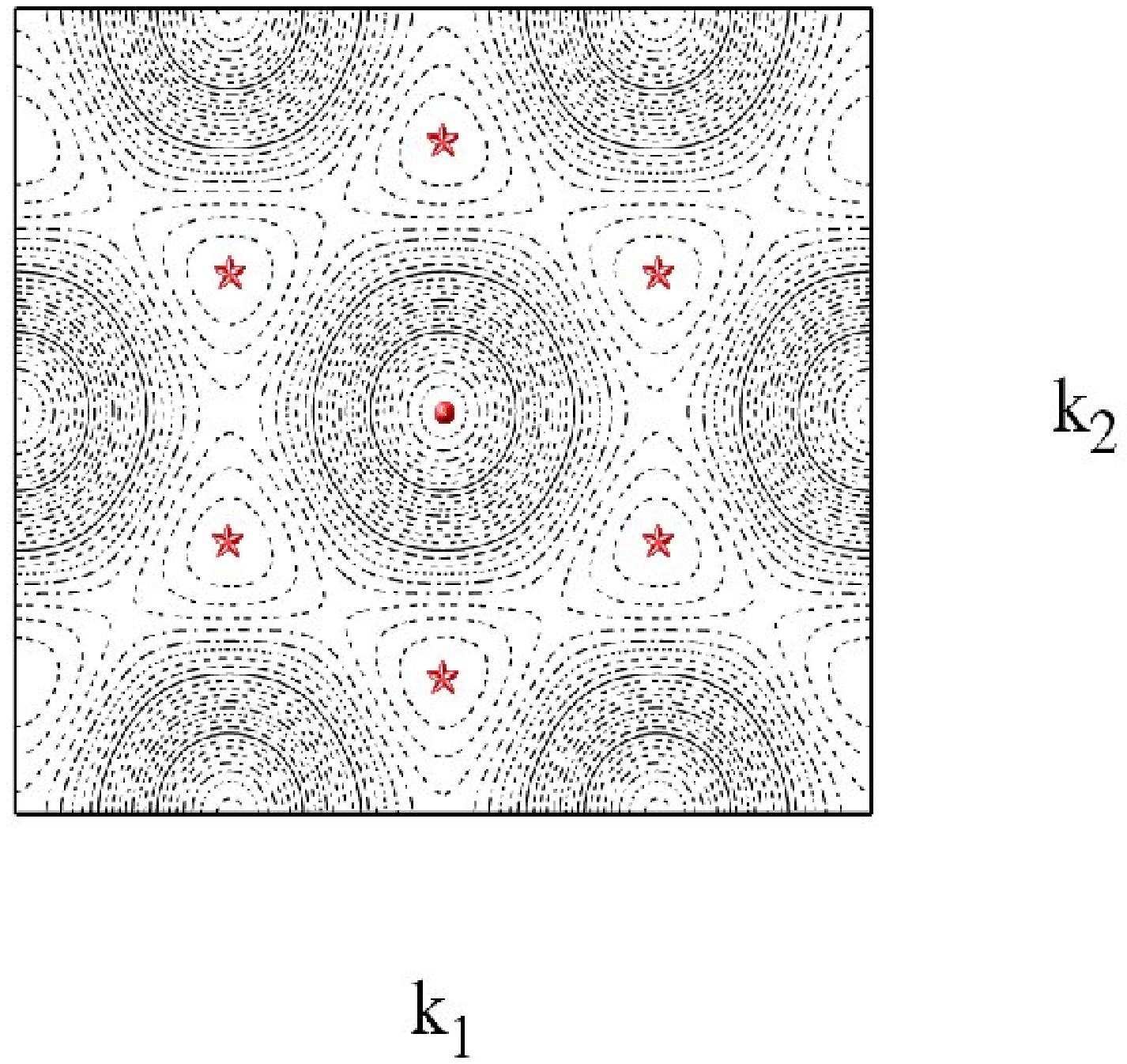,angle=-0,width=9.25cm} \vskip-2.5cm
\end{center}
\vskip-2.5cm
\caption{\it Circular hole pockets on the honeycomb lattice. 
The dot corresponds to the point Gamma and the stars mark the 
centers of the hole pockets (corresponding to the point W and its symmetry 
partners). The parameters
are the same as in figure \ref{single_landscape}.}
\label{circular_pocket}
\end{figure}

Motivated by baryon chiral perturbation theory for QCD 
\cite{Geo84a,Gas88,Jen91,Ber92,Bec99}, based on symmetry considerations,
a universal effective theory for magnons and charge carriers in 
lightly doped antiferromagnets on the square lattice
has been constructed using  
the known location of hole or electron pockets. This powerful method was
used to systematically construct the effective theory for $t$-$J$-type 
models on the square lattice in 
\cite{Kampfer:2005ba,Brugger:2006dz,Brugger:2006ej}. The effective 
theories were used to investigate the one-magnon exchange 
potential and the resulting bound states between two holes and two electrons 
as well as the possible existence of spiral phases \cite{Brugger:2006ej,
Brugger:2006dz, Brugger:2006fi}.
Using the information about the location of the pockets and based on the 
symmetry properties of the underlying 
microscopic theory, we have constructed a systematic low-energy effective 
theory for the $t$-$J$ model on the honeycomb lattice. The details of the
construction of the effective theory will be described in a forthcoming 
publication. Here we briefly sketch the principles
behind this construction and present the terms in the effective Lagrangian 
that are relevant to our present study. 
In the effective theory, the
holes which reside in momentum space pockets centered at
\begin{equation}
k^{\alpha}=(0,\frac{4\pi}{3\sqrt{3} a}),\,\,
k^{\beta}=(0,-\frac{4\pi}{3\sqrt{3} a})
\end{equation} 
are represented by
Grassmann fields $\psi_s^{f}(x)$. Here the ``flavor" index $f=\alpha, \beta$
characterizes the corresponding hole pocket and the index $s=\pm$ denotes spin
parallel $(+)$ or anti-parallel $(-)$ to the local staggered magnetization. 
The magnons are coupled to the holes through a nonlinear realization of the 
spontaneously broken $SU(2)_s$ symmetry. The
global $SU(2)_s$ symmetry then manifests itself as a local $U(1)_s$ symmetry
in the unbroken subgroup. This construction leads to an Abelian ``gauge''
field $v^3_{\mu}(x)$ and to two vector fields
$v^{\pm}_{\mu}(x)$ which are ``charged'' under $U(1)_s$ spin transformations.
The coupling of magnons 
and holes is realized through $v^{3}_{\mu}(x)$ and $v^{\pm}_{\mu}(x)$.
These fields have a
well-defined transformation behavior under the symmetries which the
effective theory inherits from the underlying microscopic models.
Based on symmetry considerations, we have constructed the leading order terms
of the effective Lagrangian for magnons and holes on the honeycomb lattice.
In this paper we only list
those terms that are relevant for the propagation of a single hole, i.e. 
\begin{equation}
\label{kinetic}
{\cal L}=\sum_{\ontopof{f=\alpha,\beta}{\, s = +,-}} \Big[ 
M \psi^{f\dagger}_s \psi^f_s + \psi^{f\dagger}_s D_t \psi^f_s 
+\frac{1}{2 M'} D_i \psi^{f\dagger}_s D_i \psi^f_s \Big]\,.
\end{equation}  
Here $M$ is the rest mass and $M'$ is the kinetic mass of a hole, while 
$D_{\mu}$ is a covariant derivative given by
\begin{eqnarray}
D_{\mu} \psi_\pm^{f}(x)&=&[\p_{\mu} \pm i v_{\mu}^{3}(x)] \ \psi_\pm^{f}(x)\,.
\end{eqnarray}
Eq.(\ref{kinetic}) yields circular hole pockets for small momenta 
which is indeed confirmed in figure \ref{circular_pocket}. 
The low-energy constant $M'$ in Eq.(\ref{kinetic}) is obtained from the 
curvature of the dispersion $E_{h}(k)$ near a minimum. For example, 
on a honeycomb lattice with 3456 spins and $J/t = 2.0$, we find $M'=
4.1(1)/(ta^2)$. 

\begin{figure}
\begin{center}
\epsfig{file=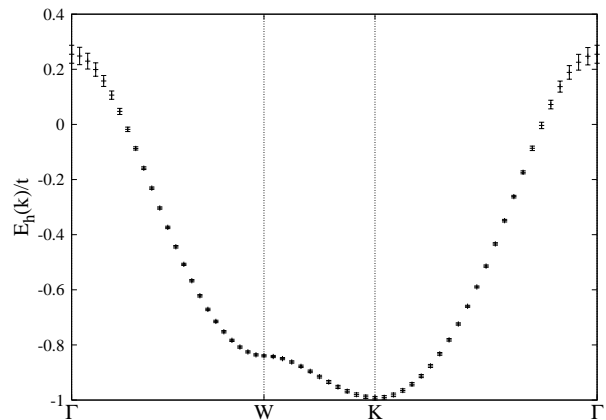,angle=0,width=8.05cm} \vskip-0.25cm
\end{center}
\caption{\it Dispersion relation $E_{h}(k)/t$ of a single hole for $J/t = 1.0$ 
along the irreducible wedge $\Gamma$-$W$-$K$-$\Gamma$ in the first Brillouin 
zone (see figure 2).}
\label{bandwidth}
\end{figure}

\begin{figure}
\begin{center}
\epsfig{file=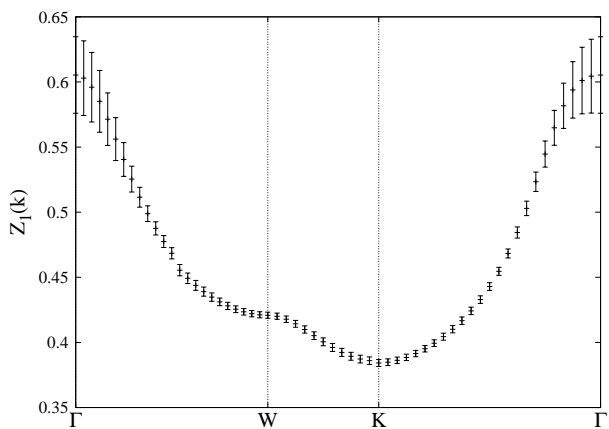,angle=0,width=8.15cm} \vskip-0.25cm
\end{center}
\caption{\it The quasiparticle weight $Z_1(k)$ of a single hole 
for $J/t = 1.0$ along the 
irreducible wedge $\Gamma$-$W$-$K$-$\Gamma$ in the first Brillouin zone (see 
figure 2).}
\label{quasiparticle}
\end{figure}

In figures \ref{bandwidth} and \ref{quasiparticle}, we have plotted the 
single-hole dispersion
as well as the quasiparticle residue $Z_1(k)$ over the irreducible wedge 
$\Gamma$-$K$-$W$-$\Gamma$ of the 
Brillouin zone for $J/t = 1.0$. The resulting bandwidth 
\begin{equation}
\Delta = E_h(\Gamma)-E_h(W)
\end{equation}
is in qualitative 
agreement with exact diagonalization and series expansion in
\cite{Luscher04}. While 
exact diagonalization of small systems may suffer from finite size effects,
and series expansions may not converge in all regions of parameter space,
the Monte Carlo data obtained with the efficient loop-cluster algorithm
do not suffer from systematic uncertainties. In table 2 
we list the kinetic mass $M'$ as well as the bandwidth $\Delta$ for a 
few values of $J/t$.  

\begin{table}
\label{data1}
\begin{center}
\begin{tabular}{|c|c|c|}
\hline 
$J/t$&
$M'ta^2$&
$\Delta /t$\tabularnewline
\hline
\hline 
$2.0$&
$4.1(1)$&
$1.15(3)$\tabularnewline
\hline 
$1.5$&
$2.9(1)$&
$1.25(3)$\tabularnewline
\hline 
$1.0$&
$1.9(1)$&
$1.24(4)$\tabularnewline
\hline 
$0.9$&
$1.8(1)$&
$1.15(6)$\tabularnewline
\hline 
$0.6$&
$1.5(2)$&
$0.9(1)$\tabularnewline
\hline
\end{tabular}
\vskip-1.25cm
\end{center}
\caption{\it Kinetic mass $M'$ as well as the bandwidth $\Delta$ for some
values of $J/t$.}
\end{table}


\section{Conclusions}
\label{concl}
We have studied antiferromagnetism on the honeycomb lattice by first principle
Monte Carlo calculations. In particular, we have fitted
more than hundred Monte Carlo data obtained at rather different volumes and 
temperatures with just four low-energy parameters --- ${\cal M}_s$, $\rho_s$,
$c$, and $M'$ --- of the effective theory for the $t$-$J$ model. These
parameters have been determined with percent and sometimes even with permille
accuracy. This should demonstrate convincingly that the systematic low-energy
effective field theory yields quantitatively correct results for the physics
of magnons and holes. Hence,
the effective theory allows us to 
perform unbiased investigations of the low-energy physics of the 
system. The construction of the effective theory for lightly doped 
antiferromagnets, as well as a systematic investigation of the one-magnon 
exchange potential and the resulting bound states 
between two holes as well as the possible existence of spiral phases of 
lightly doped antiferromagnets on the honeycomb lattice will be presented 
in subsequent studies.

\vskip-0.5cm
\section*{Acknowledgements}

We have benefited from discussions with P.\ Hasenfratz and F.\ Niedermayer. 
We also thank O.\ P.\ Sushkov for helpful correspondence. We acknowledge the 
support of the Schweizerischer Natio\-nal\-fonds.


\begin{thebibliography}{10}

\bibitem{Bednorz86} G.~J. Bednorz and K.~A. M\"uller, Z. Phys. B {\bf 64},
  188  (1986).

\bibitem{Eder96} R. Eder, Y. Ohta, and G.~A. Sawatzky, \prb {\bf 55}, R3414 
(1996).

\bibitem{Lee97} T.~K. Lee and C.~T. Shih, \prb {\bf 55}, R5983 (1997).

\bibitem{Hamer98} C.~J. Hamer, W. Zheng, and J. Oitmaa, \prb {\bf 58}, 
15508 (1998).

\bibitem{Brunner00} M. Brunner, F.~F. Assaad, and A. Muramatsu, \prb {\bf 62}
, 15480 (2000).

\bibitem{Mishchenko01} A.~S. Mishchenko, N.~V. Prokof'ev, and 
B.~V. Svistunov, \prb {\bf 64}, 033101 (2001).

\bibitem{Wel95}
B.\ O.\ Wells, Z.-X.\ Shen, D.\ M.\ King, M.\ H.\ Kastner, M.\ Greven, and
R.\ J.\ Birgeneau, Phys.\ Rev.\ Lett.\ {\bf 74}, 964 (1995).

\bibitem{LaR97}
S.\ La Rosa, I.\ Vobornik, F.\ Zwick, H.\ Berger, M.\ Grioni, G.\ Margaritondo,
R.\ J.\ Kelley, M.\ Onellion, and A.\ Chubukov, Phys.\ Rev.\ B {\bf 56}, 
R525 (1997).

\bibitem{Kim98}
C.\ Kim, P.\ J.\ White, Z.-X.\ Shen, T.\ Tohyama, Y.\ Shibata, S.\ Maekawa,
B.\ O.\ Wells, Y.\ J.\ Kim, R.\ J.\ Birgeneau, and M.\ A.\ Kastner, 
Phys.\ Rev.\ Lett.\ {\bf 80}, 4245 (1998).

\bibitem{Cha89}
S.\ Chakravarty, B.\ I.\ Halperin, and D.\ R.\ Nelson, Phys.\ Rev.\ B
{\bf 39}, 2344 (1989).

\bibitem{Neu89}
H.\ Neuberger and T.\ Ziman, Phys.\ Rev.\ B
{\bf 39}, 2608 (1989).

\bibitem{Fis89}
D.\ S.\ Fisher, Phys.\ Rev.\ B
{\bf 39}, 11783 (1989).

\bibitem{Has90}
P.\ Hasenfratz and H.\ Leutwyler, Nucl.\ Phys.\ {\bf B343}, 241 (1990).

\bibitem{Has91}
P.\ Hasenfratz and F.\ Niedermayer, Phys.\ Lett.\ {\bf B268}, 231 (1991).

\bibitem{Kampfer:2005ba}
  F.~K\"ampfer, M.~Moser, and U.-J.~Wiese,
  Nucl.\ Phys.\ {\bf B729}, 317 (2005).

\bibitem{Brugger:2006dz}
  C.~Br\"ugger, F.~K\"ampfer, M.~Moser, M.~Pepe, and U.-J.~Wiese,
  Phys.\ Rev.\  B {\bf 74}, 224432 (2006).

\bibitem{Eve93}
H.\ G.\ Evertz, G.\ Lana, and M.\ Marcu, Phys.\ Rev.\ Lett.\ {\bf 70}, 
875 (1993).

\bibitem{Wie94}
U.-J.\ Wiese and H.-P.\ Ying, Z.\ Phys.\ B {\bf 93}, 147 (1994).

\bibitem{Bea96}
B.\ B.\ Beard and U.-J.\ Wiese, Phys.\ Rev.\ Lett.\ {\bf 77}, 5130 (1996).

\bibitem{Motrunich04}
I. O. Motrunich and P. A. Lee, \prb {\bf 69}, 214516 (2004).

\bibitem{Zheng04} W. Zheng, J. Oitmaa, C.~J. Hamer, and R.~R.~P. Singh, \prb
  {\bf 70}, 020504 (2004).

\bibitem{Watanabe05} H. Watanabe and M. Ogata, J. Phys. Soc. Jpn. {\bf 74}, 
2901 (2005).

\bibitem{Net07}
A.~H.~Castro Neto, F.\ Guinea, N.~M.~R. Peres, K.~S.\ Novoselov, 
and A.~K.\ Geim, arXive:0709.1163.

\bibitem{Her06}
I.\ F.\ Herbut, Phys.\ Rev.\ Lett.\ {\bf 97}, 146401 (2006).

\bibitem{Has93}
P.\ Hasenfratz and F.\ Niedermayer, Z.\ Phys.\ B {\bf 92}, 91 (1993).


\bibitem{Mat94}
A.\ Mattsson and P.\ Fr\"ojdh, Phys.\ Rev. \ B {\bf 49}, 3997 (1994).

\bibitem{Wei91}
Z.\ Weihong, J.\ Oitmaa, and C.\ J.\ Hamer, 
Phys.\ Rev. \ B {\bf 44}, 11869 (1991).

\bibitem{Oit92}
J.\ Oitmaa, and C.\ J.\ Hamer, and Z.\ Weihong, 
Phys.\ Rev. \ B {\bf 45}, 9834 (1992).

\bibitem{Cas05}
E.\ V.\ Castro, N.\ M.\ R.\ Peres, D.\ S.\ D.\ Beach, and A. W. Sandvik,
Phys.\ Rev. \ B {\bf 73}, 054422 (2006).

\bibitem{Luscher04}
A. \ L\"uscher, A.\ L\"auchli, W.\ Zheng, and O.\ Sushkov,  
Phys.\ Rev.\ B {\bf 73}, 155118 (2006). 

\bibitem{Geo84a}
H.~Georgi, Weak Interactions and Modern Particle Theory, 
Ben\-ja\-min-Cum\-mings Publishing Company, 1984.

\bibitem{Gas88}
J.~Gasser, M.~E.~Sainio, and A.~Svarc, Nucl.\ Phys.\ 
{\bf B307}, 779 (1988).

\bibitem{Jen91}
E.~Jenkins and A.~Manohar, Phys.\ Lett.\ 
{\bf B255}, 558 (1991).

\bibitem{Ber92}
V.~Bernard, N.~Kaiser, J.~Kambor, and U.-G.~Meissner, Nucl.\ Phys.\ 
{\bf B388}, 315 (1992).

\bibitem{Bec99}
T.~Becher and H.~Leutwyler, Eur.\ Phys.\ J.\ C {\bf 9}, 643 (1999).

\bibitem{Brugger:2006ej}
  C.~Br\"ugger, C.~P.~Hofmann, F.~K\"ampfer, M.~Moser, M.~Pepe, and U.-J.~Wiese,
  Phys.\ Rev.\  B {\bf 75}, 214405 (2007).

\bibitem{Brugger:2006fi}
  C.~Br\"ugger, C.~P.~Hofmann, F.~K\"ampfer, M.~Pepe, and U.-J.~Wiese,
  Phys.\ Rev.\  B {\bf 75}, 014421 (2007).


\end{thebibliography}
\end{document}